\documentclass[aps,prb,twocolumn,showpacs,dvipsnames,superscriptaddress]{revtex4-2}

\usepackage[utf8]{inputenc}

\usepackage{amssymb,amsfonts,amsmath} 
\usepackage{graphicx,epsfig,psfrag}
\usepackage{color}
\usepackage{natbib}
\usepackage{url}
\usepackage[breaklinks=true]{hyperref}
\usepackage{mathtools}
\usepackage{subfigure}
\usepackage{physics}
\usepackage{lipsum}
\usepackage{comment}
\hypersetup{
        colorlinks = true,
        citecolor = blue
}

\begin{document}

\title{Tomasch Oscillations as Above-Gap Signature of Topological Superconductivity}

\affiliation{National Laboratory of Solid State Microstructures, School of Physics, and Collaborative Innovation Center of Advanced Microstructures, Nanjing University, Nanjing 210093, China}
\affiliation{TCM Group, Cavendish Laboratory, J. J. Thomson Avenue, Cambridge CB3 0HE, United Kingdom}
\affiliation{School of Microelectronics and Physics, Hunan University of Technology and Business, Changsha, Hunan Province, 410205, China}
\affiliation{Department of Physics, University of Konstanz, 78464 Konstanz, Germany}

\author{Antonio \v{S}trkalj}
\thanks{The authors contribute equally.}
\affiliation{TCM Group, Cavendish Laboratory, J. J. Thomson Avenue, Cambridge CB3 0HE, United Kingdom}
\author{Xi-Rong Chen}
\thanks{The authors contribute equally.}
\affiliation{National Laboratory of Solid State Microstructures, School of Physics, and Collaborative Innovation Center of Advanced Microstructures, Nanjing University, Nanjing 210093, China}
\affiliation{School of Microelectronics and Physics, Hunan University of Technology and Business, Changsha, Hunan Province, 410205, China}
\author{Wei Chen}
\email{Corresponding author: pchenweis@gmail.com}
\affiliation{National Laboratory of Solid State Microstructures, School of Physics, and Collaborative Innovation Center of Advanced Microstructures, Nanjing University, Nanjing 210093, China}
\author{D. Y. Xing}
\affiliation{National Laboratory of Solid State Microstructures, School of Physics, and Collaborative Innovation Center of Advanced Microstructures, Nanjing University, Nanjing 210093, China}
\author{Oded Zilberberg}
\affiliation{Department of Physics, University of Konstanz, 78464 Konstanz, Germany}

\begin{abstract}
The identification of topological superconductors usually involves searching for in-gap modes that are protected by topology. However, in  current experimental settings, the smoking-gun evidence of these in-gap modes is still lacking. 
In this work, we propose to support the distinction between two-dimensional conventional $s$-wave and topological $p$-wave superconductors by above-gap transport signatures. Our method utilizes the emergence of Tomasch oscillations of quasiparticles in a junction consisting of a superconductor sandwiched between two metallic leads. 
We demonstrate that the behavior of the oscillations in conductance as a function of the interface barriers provides a distinctive signature for $s$-wave and $p$-wave superconductors. Specifically, the oscillations become weaker as the barrier strength increases in $s$-wave superconductors, while they become more pronounced in $p$-wave superconductors, which we prove to be a direct manifestation of the pairing symmetries. 
Our method can serve as a complimentary probe for identifying some classes of topological superconductors through the above-gap transport.
\end{abstract}
\maketitle

%
%
\textit{Introduction.--}  At the heart of superconductivity is the pairing of conduction electrons into Cooper pairs that form a bosonic condensate~\cite{Bardeen1957}. These Cooper pairs can be in either a spin-singlet state, with a total spin 0, or a spin-triplet state, with a spin 1. The spin-singlet state is characterized by a wavefunction with even angular momentum, such as $s$-wave or $d$-wave, while the spin-triplet state supports a wavefunction with odd angular momentum, such as $p$-wave or $f$-wave.
In conventional $s$-wave superconductivity the pairing function  $\Delta(\mathbf{k}) = \Delta_s$ is constant irrespective of the direction of the momentum vector $\mathbf{k}$. The Cooper pair, in this case, consists of two electrons with opposite spins. On the other hand, in unconventional $p$-wave superconductors, electrons with the same spin form Cooper pairs and the pairing $\Delta(\mathbf{k})$ is no longer constant with $\mathbf{k}$~\cite{Sigrist1991,Mackenzie2003,Hasan2010,Qi2011,Tanaka2012,Sato2017}.

Over the past several decades, there has been significant interest in unconventional $p$-wave superconductors~\cite{Bednorz1986,Sigrist1991,Mackenzie2003,Hasan2010,Qi2011,Sato2017}, mostly due to their unique topological properties~\cite{Fu2008,Leijnse2012,Bernevig2013}. 
The topology implies, for example, the presence of exotic quasiparticles, such as Majorana zero-energy modes in one-dimensional systems~\cite{Kitaev2001,Lobos2015}, and in-gap Majorana states in two-dimensional systems~\cite{Volovik1999,Read2000,Ivanov2001,Biswas2013,Sun2016,Li2016,Brun2016}. These quasiparticles have potential applications in topological quantum computing~\cite{Kitaev2003,Nayak2008,Alicea2011}, and are used as a key signature for discerning between topologically nontrivial $p$-wave and trivial $s$-wave superconductors. As a result, the search for topological superconductors encompass two main approaches. The first approach involves searching for topological superconductivity in specific materials, such as Sr$_2$RuO$_4$~\cite{Rice1995,Baskaran1996,Mackenzie2003,Sigrist2005}, UTe$_2$~\cite{Jiao2020}, Pb$_3$Bi~\cite{Qin2023}, and hybrid systems such as Pb/Co/Si(111)~\cite{Menard2017,Brun2016}. The second approach involves using engineered metamaterials that share some of the properties of topological superconductors~\cite{Lutchyn2010,Oreg10prl,Klinovaja2013,Beenakker2016,Lado2018,Volpez2019,Fu2008,Sato2009,Sau2010,Alicea2010,Mourik2012,Das2012,Albrecht2016,Li2016}. 

The behavior of superconductors can be largely understood through the Bogoliubov-de-Gennes (BdG) formalism~\cite{deGennes1966}, which describes the mean-field behavior of quasiparticles in the superconductor
that form through hybridization between electrons and holes. The resulting band structure resembles a "sombrero" shape, with an energy gap that depends on the details of the pairing function $\Delta(\mathbf{k})$. 
Notably, the band structure of quasiparticles in superconductors resembles that of band-inverted semiconductors~\cite{Altarelli1983,Yang1997,Lakrimi1997,Karalic2020}, with the band gap of the latter corresponding to the superconducting gap in the former. 
Recently, Fabry-Pérot oscillations were observed in a two-dimensional junction made out of an inverted InAs/GaSb double quantum well~\cite{Karalic2020}. 
The mechanism leading to such Fabry-Pérot oscillations in a two-dimensional junction stems from the sombrero-shaped band structure. Specifically, the interference is dominated by the scattering between the electron-like and hole-like states at energies close to the band gap~\cite{Karalic2020,Zhao2023}.
Such interference is quite ubiquitous and applies to a variety of condensed matter systems with inverted-band dispersions~\cite{Karalic2020,Li2022}. Interestingly, similar Fabry-Pérot oscillations are also studied in superconducting junctions and go under the name of Tomasch oscillations~\cite{Tomasch1965,Tomasch1966,McMillan1966,McMillan1968,Granqvist1974,Nedellec1976,Kononov2018}.

In this work, we demonstrate that Tomasch oscillations in the conductance across the two-dimensional NSN junctions (two normal metals sandwiching a superconductor) provide a signature that can act as a complementary probe to discern between conventional and topological superconductors.
As such, we focus on transport with energies above the superconducting gap and investigate the effects of the interface barriers on the Tomasch oscillations for superconductors with different pairing symmetries. In the weak barrier limit, we find that the inverted-band mechanism responsible for the oscillations is the same for both $s$-wave and $p$-wave superconductors. Interestingly, the oscillations are crucially different in the strong barrier limit, i.e., in the tunneling limit. This distinction arises from differing pairings in the BdG Hamiltonians of $s$- and $p$-wave superconductors, affecting the visibility of the oscillations. Our result offers an alternative experimental probe using the \textit{above-gap} transport signatures for distinguishing between conventional and topological superconductors, in contrast to the commonly studied in-gap signatures~\cite{Kokkeler2022}.

%
%
\begin{figure}[t!]
	\centering
    \includegraphics[scale=1]{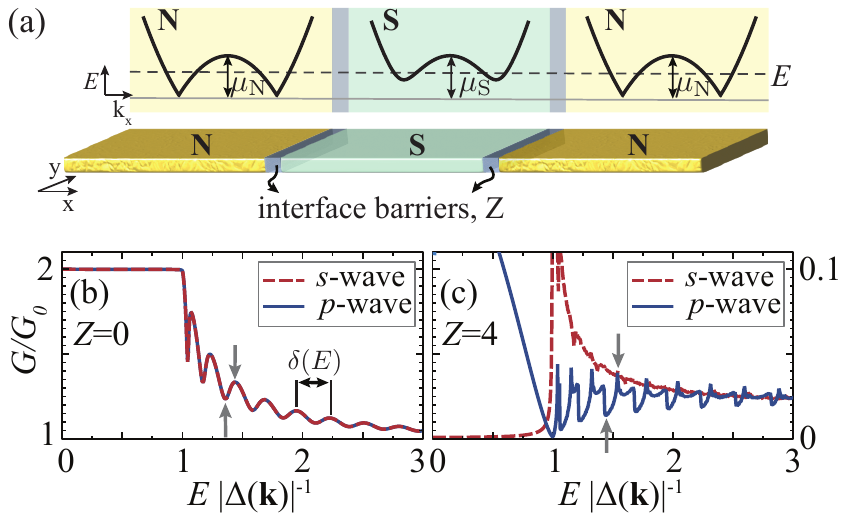}
    \caption{
    (a) Sketch of the NSN junction consisting of a superconductor (S) coupled to two normal metal leads (N) with the interface barrier $Z$. (top panels) The quasiparticle dispersion for $k_y=0$ in all three regions. We consider an incident particle from the left lead with energy $E$ (dashed line). 
    (b) Differential conductance as a function of energy for both the $s$- and $p$-wave superconductors in the absence of barriers ($Z=0$) and with the magnitude of the pair potential set to be equal, \textit{i.e.}, $|\Delta(\mathbf{k})| = \Delta_s=\Delta_p k_{\rm F}$. The period of the conductance oscillations, $\delta (E)$, increases with energy. (c) Same as (b) in the presence of finite barriers with equal strengths $Z=4$.  
    Small grey arrows denote the minimum and the maximum of a single oscillation that we use in Fig.~\ref{fig:figure3} when calculating the averaged visibility.
    For (b) and (c), we used the following parameters: $L=10 \xi_0$, $k_{\rm F}=2000 \xi_0^{-1}$, $\mu_{S}=\mu_{\rm N}=1000 |\Delta(\mathbf{k})|$, where $|\Delta(\mathbf{k})|$ is the magnitude of the pair potential and $\xi_0$ is the superconducting coherence length.
    }
    \label{fig:figure1}
\end{figure}

\textit{Setup.--}  
We study a two-dimensional NSN junction made of two normal metals (N), and a superconductor (S) sandwiched between them, see Fig.~\ref{fig:figure1}(a).  We concentrate on a ballistic case, where the mean free path of particles is the largest scale in the system.
The BdG Hamiltonian around $\Gamma$ point in the continuum limit
of the whole system is
\begin{align}  \label{eq:hamiltonian}
    H = H_{\rm N} + H_{\rm S} + U(x) \sigma_z\, ,
\end{align}
where the Pauli matrix $\sigma_z$ (and later $\sigma_x$) operates in the Nambu space and $H_{\rm N}$ describes the metallic leads
\begin{align}  \label{eq:hamiltonian_N}
    H_{\rm N}(\mathbf{k}) &=
        \begin{pmatrix}
        \frac{\hbar^2}{2 m_{\rm N}} \mathbf{k}^2 - \mu_{\rm N} & 0\\
        0 & -\frac{\hbar^2}{2 m_{\rm N}} \mathbf{k}^2 + \mu_{\rm N}
        \end{pmatrix} \,,
\end{align}
with $\mathbf{k}=(k_x, k_y)$ the wave vector, $m_{\rm N}$ the effective mass of electrons in the metallic leads, and $\mu_{\rm N}$ their chemical potential. The Hamiltonian of the superconductor is given by~\cite{footnote_basis}
\begin{align}  \label{eq:hamiltonian_S}
    H_{\rm S}(\mathbf{k}) &=
        \begin{pmatrix}
        \frac{\hbar^2}{2 m_{\rm S}} \mathbf{k}^2 - \mu_{\rm S} & \Delta(\mathbf{k})\\
        \Delta^*(\mathbf{k}) & -\frac{\hbar^2}{2 m_{\rm S}} \mathbf{k}^2 + \mu_{\rm S}
        \end{pmatrix} \, ,
\end{align}
with $m_{\rm S}$ and $\mu_{\rm S}$ the corresponding effective mass and chemical potential, respectively. We introduce a pairing potential $\Delta(\mathbf{k})$ corresponding to two types of superconductors, namely a time-reversal symmetric $s$-wave superconductor with a constant pairing $\Delta(\mathbf{k}) = \Delta_s$, and a time-reversal broken $p$-wave superconductor with $\Delta(\mathbf{k}) = i\Delta_p (k_x + i k_y)$. 
Note that for a vanishing pairing potential, the band structure takes a parabolic shape. Furthermore, at the N-S interfaces, we introduce sharp barriers $U(x) = U (\delta(x) + \delta(x-L))$ to account for the materials' mismatch or imperfections~\cite{Blonder1982,Scheer2001}. Alternatively, the barriers can be introduced and adjusted by local strip gates. For simplicity, we assume that the barriers are perfectly flat in the $y$-direction, such that they do not break the translational invariance in the $y$-direction. In the following, we employ the often-used dimensionless barrier strength $Z \equiv m U / (\hbar^2 k_{\rm F})$ and, without loss of generality consider $m\equiv m_{\rm N}=m_{\rm S}$ and $\mu\equiv \mu_{\rm N}=\mu_{\rm S}=\hbar^2 k_{\rm F}^2/(2m)$.

We calculate the differential conductance across the junction using the Blonder-Tinkham-Klapwijk formula~\cite{Blonder1982}
{\small
\begin{equation}  \label{eq:conductance}
    G(E) = G_0 \int^{K(E)}_{-K(E)} \frac{\dd k_y}{2K(E)} \left[ 1+|a_L(k_y,E)|^2-|b_L(k_y,E)|^2 \right] \,,
\end{equation}
} 
where $a_{L}$ and $b_{L}$ denote the amplitudes of Andreev~\cite{Andreev1964} and normal reflections, respectively. The factor $K(E)=\sqrt{2 m_{\rm N} (E+\mu)}$ is the maximal value of $k_y$ for a given incident energy $E$, and $G_0=2e^2KW/(\pi h)$ is the conductance of the metallic lead with width $W$ in the $y$-direction. The microscopic analysis of the scattering amplitudes appears below, cf.~Eq.~\eqref{eq:scattering_eqs}. Using formula~\eqref{eq:conductance}, we calculate the conductance for $s$- and $p$-wave superconductors for vanishing barrier ($Z=0$) and strong barrier ($Z=4$), see Figs.~\ref{fig:figure1}(b) and (c), respectively.
The case of a vanishing/weak barrier shows the same conductance behavior for both the $s$- and $p$-wave case, namely strong oscillations with an energy-dependent period $\delta(E)$. The oscillations come from constructive interference of multiple scattering paths inside the superconducting cavity, where electronic modes in the outer branch of the band structure scatter to hole modes in the inner part of the band structure. Therefore, it is possible to analytically obtain the position of each peak in the conductance, as well as their period, by solving the following interference condition $k_x^{e/h}(E)-k_x^{h/e}(E) = 2 \pi n / L$, where $n$ is an integer. This limit of vanishing barriers is studied in detail in Refs.~\cite{Karalic2020,thesis_Strkalj}.
On the other hand, in the opposite limit of strong barriers [cf.~the $Z=4$ case in Fig.~\ref{fig:figure1}(c)], the conductance oscillations drastically differ between the $s$- and $p$-wave cases. Specifically, they are suppressed in the former and significantly enhanced in the latter. This signature in bulk transport above the gap is the main result of our work and it can be used to distinguish between different types of superconductors.

\begin{figure}[t]
	\centering
    \includegraphics[scale=1]{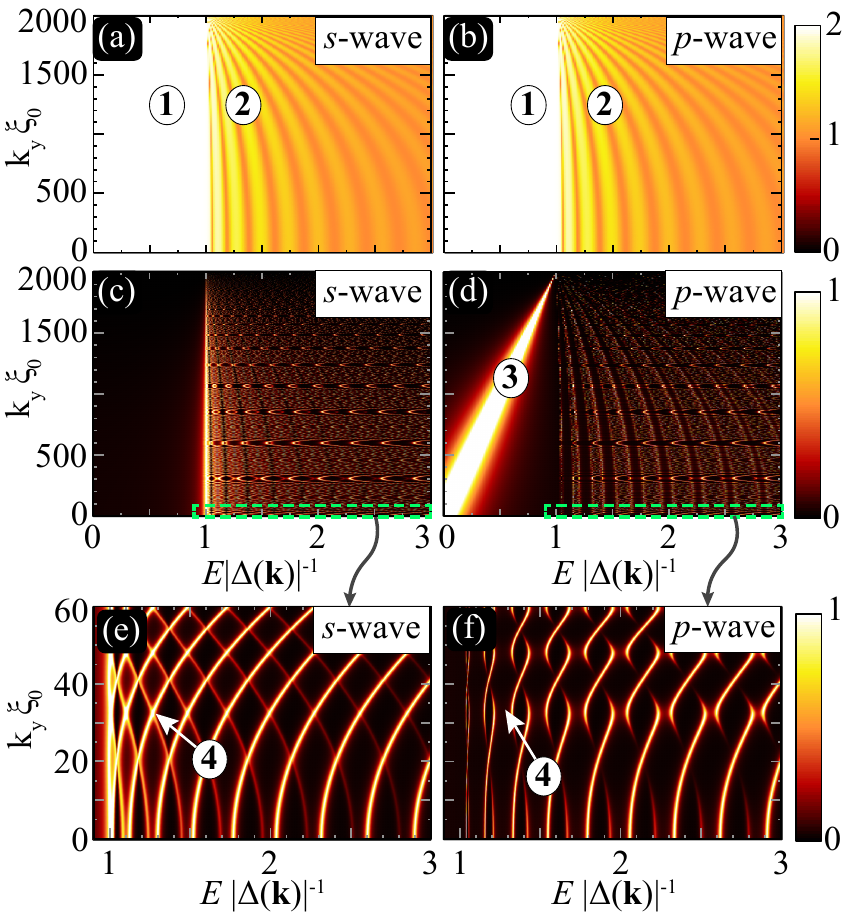}
    \caption{
    (a) and (b) Transmission probability with $Z=0$ for $s$- and $p$-wave superconductor, respectively.
    Two main features are visible: \textcircled{1} inside the gap, perfect transmission occurs with $T=2$, and \textcircled{2} both electron- and hole-like quasiparticles are present, hybridize with one another, and produce oscillations that only weakly depend on $k_y$.
    (c) and (d) Transmission probability with $Z=2$ for $s$- and $p$-wave superconductor, respectively. In (d) the topological in-gap mode, marked with \textcircled{3}, is visible, while in (c) there are no in-gap modes present.
    (e) and (f) The enlarged regions of (c) and (d). A new differentiating feature, marked with \textcircled{4}, appears in a case of strong barriers: in (e), the transmission maxima of different electron- and hole-like modes cross, while in (f) avoided crossings appear, leading to secondary gaps and flat transmission bands appear.
    For all plots, we used the same parameters as in Fig.~\ref{fig:figure1}.
    }
    \label{fig:figure2}
\end{figure}

\textit{Momentum resolved transmission.--}
To understand the microscopic origin of the results shown in Figs.~\ref{fig:figure1}(b) and (c), we turn to the analysis of the momentum-resolved transmission probability $T(k_y,E)=1+|a_L(k_y,E)|^2-|b_L(k_y,E)|^2$, \textit{i.e.}, the integrand of Eq.~\eqref{eq:conductance},
which describes the transmission of charge in the left lead.
We start by formulating the junction's scattering equations by considering an electron incident from the left lead with energy $E$ and transverse momentum $k_y$. We describe the states in the left/right metallic leads (L/R) and the superconducting region (S)

\begin{equation}  \label{eq:scattering_eqs}
\begin{split}
  \Psi_{L} = &\left[ \vec{\Phi}_N^e e^{i q_x^e x} + a_{L} \vec{\Phi}_N^h e^{i q_x^h x}  + b_{L} \vec{\Phi}_N^e e^{-i q_x^e x}
  \right] e^{i k_y y},   \\
  \Psi_{R} = &\left[ a_{R} \vec{\Phi}_N^h e^{-i q_x^h x}  + b_{R} \vec{\Phi}_N^e e^{i q_x^e x}
  \right] e^{i k_y y},  \\
  \Psi_{S} = &\sum_{\eta=\pm}\left[ s^e_{\eta} \vec{\Phi}_{S}^e  e^{i \eta k^e_x x}
  + s^h_{\eta} \vec{\Phi}_{S}^h e^{-i \eta k_x^h x}
  \right] e^{i k_y y},
  \end{split}
\end{equation}
where $q_x^{e/h}$ and $k_x^{e/h}$ are the $x$-components of the quasiparticle's momentum in the metallic leads and the superconductor, respectively. In the metallic leads, the spinors
$\vec{\Phi}_N^{e}=(1,0)^T, \vec{\Phi}_N^{h}=(0,1)^T$ describe an electron in the outer dispersion branch and a hole in the inner dispersion branch, respectively. 
$\vec{\Phi}_{S}^{e} = [u(k^e_x,k_y), v(k^e_x,k_y)]^T$ and $\vec{\Phi}_{S}^{h} = [u(k^h_x,k_y), v(k^h_x,k_y)]^T$ are spinors of electron- and hole-like quasiparticles in the superconductor and $u,v$ are electron and hole wave components. 
The coefficients $a_{L}, a_{R}, b_{L}, b_{R}$ denote the amplitudes of Andreev reflection~\cite{Andreev1964}, cross Andreev reflection, normal reflection, and elastic cotunneling, respectively. The coefficients $s^{e,h}_{\pm}$ are scattering amplitudes inside the superconductor. 

To find the scattering amplitudes above, we impose the following boundary conditions on the two N-S interfaces: 
$\Psi_{L/R}=\Psi_{S}$ and $\partial_x\Psi_{S}-\partial_x\Psi_{L/R}=\pm 2 Z k_F \Psi_{S}$ for the $s$-wave superconductor, and $\Psi_{L/R}=\Psi_S$ and $\partial_x\Psi_S-\partial_x\Psi_{L/R}= \pm 2 Z k_F \Psi_{S} + (m \Delta/\hbar) \sigma_x \Psi_{S}$ for the $p$-wave superconductor, where ``$\pm$" correspond to the left and right interfaces at $x=0,L$, respectively. 
Note that due to the perfectly flat barriers in $y$-direction, the momentum $k_y$ is preserved for all scattering processes.

We solve the scattering equations~\eqref{eq:scattering_eqs}, with the aforementioned boundary conditions, for $a_L$ and $b_L$, and show the result for $T(k_y,E)$ in Fig.~\ref{fig:figure2} for both the $s$- and $p$-wave superconductors and in the limits of weak [Figs.~\ref{fig:figure2}(a) and (b)] and strong [Figs.~\ref{fig:figure2}(c-f)] barriers. 
In the former, $T(k_y,E)$ is identical for $s$- and $p$-wave superconductors and we identify two main features (marked \textcircled{1} and \textcircled{2} in the figure). 
In region \textcircled{1}, the energy resides in the superconducting gap, i.e., the \textit{main gap}, see Fig.~\ref{fig:figure1}(a) and (b), and the transmission through the NSN junction is constant and equal to 2 without the effect of the interface barriers~\cite{Andreev1964, Blonder1982}. 
In region \textcircled{2}, both electron and hole-like modes coexist; due to the predominant electron-to-hole scattering, the transmission maxima exhibit a relatively weak dependence on $k_y$. Consequently, strong oscillations manifest in the conductance, see Fig.~\ref{fig:figure1}(b).
Note that $T(k_y,E) = T(-k_y,E)$, and therefore in Fig.~\ref{fig:figure2}, we show only positive $k_y$ plane. 

In the opposite limit of strong barriers -- or equivalently, weak coupling to the leads --  the transmission throughout region \textcircled{1} is now strongly suppressed for the $s$-wave superconductor, with a power-law scaling with $Z$~\cite{Blonder1982}. At the same time, for the $p$-wave superconductor on top of the suppressed transmission, a clear sign of a topological edge mode, marked with \textcircled{3}, can be seen~\cite{Bernevig2013}. 
On the other hand, the maxima of $T(k_y,E)$ in region \textcircled{2} become sharper for both $s$-wave and $p$-wave superconductor and they also acquire an additional structure that was smeared out by the strong coupling with the leads, see Figs.~\ref{fig:figure2}(c-f). Moreover, gaps between maxima -- dubbed \textit{secondary gaps} -- close in the case of the $s$-wave superconductor, while in the $p$-wave case they remain open even for very large barrier strengths $Z$, as marked by \textcircled{4} in Figs.~\ref{fig:figure2}(e) and (f).

\textit{Opening of secondary gaps.--}
To better understand the mechanisms responsible for the different behavior of the secondary gaps in transmission -- and with that the difference in conductance oscillations -- between the $s$- and $p$-wave superconductors, we employ a perturbative approach.
We first concentrate on the limit of strong barriers.
In this limit, the gap structure in the transmission plots [cf.~Figs.~\ref{fig:figure2}(c-f)] is determined by the eigenmodes of the superconducting cavity, which are only perturbatively affected by the coupling to the leads.
Therefore, it is sufficient to analyze the isolated superconductor, which is finite in the $x$-direction with length $L$ and infinite in the $y$-direction. Doing so, we discover that the momentum dependence of the pairing potential in Eq.~\eqref{eq:hamiltonian_S} is responsible for the selective hybridization of particle- and hole-like modes of the cavity, see Supplemental Material~\cite{SM} for more details. Crucially, in the $p$-wave superconductor, secondary gaps are opened even without the presence of the leads due to the $k_x$ dependence of the pairing term in $H_{\rm S}$. The Hamiltonian of the $s$-wave superconductor, on the other hand, has constant off-diagonal elements and the particle- and hole-like modes do not hybridize. As a result, the secondary gaps in  $T(k_y,E)$ are closed in that case.

In the limit of weak barriers, i.e. the strong hybridization with the leads, secondary gaps are opened for both $s$-wave and $p$-wave superconductors, see Figs.~\ref{fig:figure2}(a) and (b).
To understand this, we include the leads in our analytical analysis via the weak tunnel coupling to the superconductor. 
Such treatment, which relies on a calculation of the self-energy, gives rise to the finite coupling between the electron- and hole-like modes of the superconductor, which is second-order in the tunneling. As a result, secondary gaps are opened between all degenerate modes of the closed cavity, see Supplemental Material~\cite{SM} for details. This conclusion is also valid when the tunnel coupling is strong, i.e., when there are no barriers at all, cf.~Figs.~\ref{fig:figure2}(a) and (b).

\begin{figure}[t!]
    \centering
    \includegraphics[scale=1]{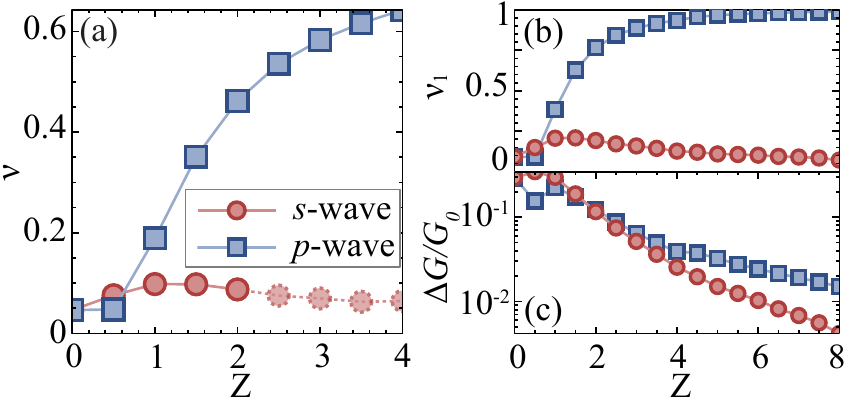}
    \caption{ 
    (a) Visibility of the conductance oscillations defined in Eq.~\eqref{eq:visibility} as a function of the barrier strength for the $s$-wave (red dots) and the $p$-wave (blue squares) cases.  
    (b) Visibility of the first oscillation, i.e., the one closest to the gap for both $s$-and $p$-wave cases. 
    (c) The height of the first oscillation, $\Delta G \equiv G_1^{\max}-G_1^{\min}$, for both cases. 
    We used the same parameters as in Fig.~\ref{fig:figure1}.
    }
    \label{fig:figure3}
\end{figure}

To quantify the impact of the barrier in both $s$- and $p$-wave cases, we study the visibility of conductance oscillations defined as 
\begin{equation}
    \nu=\frac{1}{N}\sum_{i=1}^{N}\frac{G_{i}^{\text{max}}-G_i^{\text{min}}}{G_i^{\text{max}}+G_i^{\text{min}}} \, ,
    \label{eq:visibility}
\end{equation}
where $G_i^{\text{max}}$ and $G_i^{\text{min}}$ are the neighboring local maximum and minimum values of the conductance, see grey arrows in Figs.~\ref{fig:figure1}(b) and (c). We numerically calculate the visibility of the first five periods of oscillation ($N=5$) as a function of the barrier strength $Z$, see Fig.~\ref{fig:figure3}. While in the absence of the barriers, $Z=0$, both types of superconductors have the same values of visibility, at large $Z$ the visibility increases with $Z$ for the $p$-wave superconductor and saturates to a small constant for the $s$-wave superconductor. 
Such behavior reflects the analytical discussion above on secondary gaps.
Note that when $Z > 2$ in the $s$-wave case [dashed lines inFig.~\ref{fig:figure3}(a)], the height of the conductance oscillations is so small that it becomes comparable to fluctuations caused by the finite-element numerical integration over Eq.~\eqref{eq:conductance}. On the other hand, the height of the first oscillation is distinguishable for much larger $Z$ in both the $s$- and $p$-wave cases. We plot the visibility of that first oscillation peak in Fig.~\ref{fig:figure3}(b), from which the same trend can be extracted as in Fig.~\ref{fig:figure3}(a). Lastly, in Fig.~\ref{fig:figure3}(c), we plot the height of the aforementioned first oscillation -- defined as $\Delta G \equiv G_1^{\max}-G_1^{\min}$ -- for both cases. The height decays for large $Z$, but with a slower rate in the $p$-wave case.

In conclusion, we have demonstrated that Tomasch oscillations of Bogoliubov quasiparticles provide a promising method to distinguish between topological $p$-wave superconductivity and conventional $s$-wave superconductivity. Specifically, the resulting conductance oscillations display contrasting behavior for the two types of superconductors as the interface barriers increase, which is a direct manifestation of the pairing symmetries. Our study introduces bulk probes for identifying topological superconductivity, which is usually overlooked. Our proposed above-gap transport signature can serve as an essential supplement to in-gap measurements~\cite{SM}. Interestingly, Tomasch oscillations were first reported approximately 60 years ago in both Pb and In films with thicknesses ranging from 3 to 30 $\mu$m~\cite{Tomasch1965,Tomasch1966}, making their observation in junctions with topological superconductors highly promising using state-of-the-art techniques.
Furthermore, some recent experiments on normal-metal/insulator/superconductor junctions~\cite{Cao2022} reported high tunability of the barrier strength $Z$. By changing the thickness of the barrier, $Z$ is easily tuned to $Z \gg 10$, which lies deeply in the regime that we discuss in our work.
Last, our mechanism is established for continuum single-band models without the presence of spin-orbit coupling, where the band inversion and pairing mechanism dictate the appearance of topology; it would be interesting to extend the discussion to more complicated multi-band systems as well as lattice models with anisotropy.
There, anisotropy can cause a significant morphing of the sombrero-shaped band structure, which can even lead to topological phase transitions by gap closing at momenta away from the $\Gamma$ point~\cite{Sato2010}. Whether our method can be applied to distinguish between different phases in that case will be the focus of future work.

%
\section*{Acknowledgements}
%
We are grateful to W. Belzig for insightful discussions. 
W.C. acknowledges financial support from the National Natural Science Foundation of
China under Grant No. 12074172 and No. 12222406, and the National Key Projects
for Research and Development of China under Grant No.
2022YFA120470. 
X.R.C. acknowledges financial support from the National Natural Science Foundation of
China under Grant No. 12304064.
D.Y.X. acknowledges financial support from the
State Key Program for Basic Researches of China under Grants No. 2021YFA1400403.
A.Š. acknowledges financial support from the Swiss National Science Foundation (Grant No.~199969). O.Z. acknowledges financial support from the Deutsche Forschungsgemeinschaft (DFG) - project number 449653034. All data that support the plots within this paper are available from the corresponding author upon request.

%
%
%

\newpage
$\phantomsection$
\newpage
\onecolumngrid
\renewcommand{\theequation}{S.\arabic{equation}}
\setcounter{equation}{0}
\renewcommand{\thefigure}{S.\arabic{figure}}
\setcounter{figure}{0}
%
\section*{Supplemental Material for “Tomasch Oscillations as Above-Gap Signature of Topological Superconductivity”}
%

The Supplemental Material is structured as follows: in Sec.~\ref{app_sec:isolated_cavity}, we first elucidate the mechanism of opening of secondary gaps for the $p$-wave case in an isolated superconducting cavity (SC) and the lack of it in the $s$-wave case. This explains the differences in transmission patterns (Fig.~2 in the main text) and conductance oscillations [Fig.~1(c) in the main text] in the strong barrier limit ($Z\gg1$).
In Sec.~\ref{app_sec:coupling_to_leads}, we show how coupling to the metallic leads provides the mechanism for opening secondary gaps also in the $s$-wave case.
In Sec.~\ref{app_sec:different_length}, we show the dependence of the oscillations in conductance on the length of the SC. Lastly, in Sec.~\ref{app_sec:advanteges_drawbacks}, we briefly discuss some clear advantages and potential drawbacks of our method. 

%
\subsection{Opening of secondary gaps in an isolated semi-infinite superconducting cavity  \label{app_sec:isolated_cavity}}
%
\begin{figure}[h!]
	\centering
	\includegraphics[width=0.6\columnwidth]{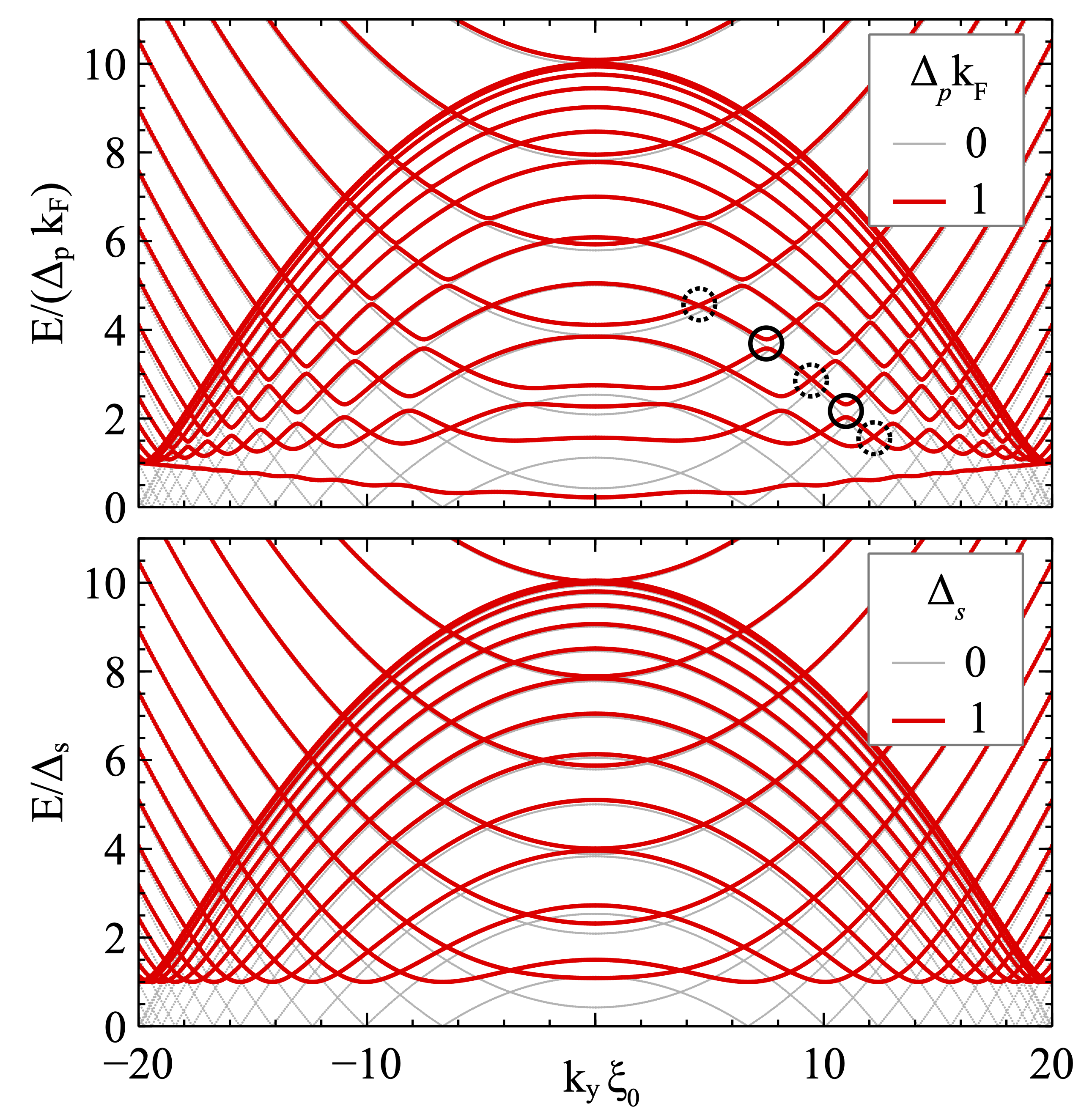}
	\caption{ 
		Band structure for an isolated superconducting cavity that is infinite in $y$-direction, and finite in $x$-direction with width $L$. The upper plot shows the case of a $p$-wave superconductor where the secondary gaps open between electronic and hole modes with different parities. Full black circles mark secondary gaps that open, while dashed circles mark the ones that remain closed even for finite $\Delta_p$.
		The lower plot shows the case of an $s$-wave superconductor where the secondary gaps remain closed for any value of $\Delta_s$. To emphasize all the features of the band structure, i.e. shape of the sombrero hat, the main gap, regimes inside/outside the sombrero hat where only particles/both particles and holes exist and secondary gaps, we changed the parameters -- compared to the main text -- and used the following: $L=2 \xi_0$, $k_{\rm F}=20 \xi_0^{-1}$, $\mu_{S}=10 \Delta_s$ ($\mu_{S}=10 \Delta_p k_F$ for the $p$-wave SC), where $\xi_0$ is the superconducting coherence length.
        Note that due to the relatively small $L$, the two in-gap modes in the upper plot are hybridised around $k_y \xi_0 = 0$, which results in a small gap opening and vanishing of the dispersion at zero energy. This effect is not observed for much larger cavities discussed in the main text, see Fig.~2(d). 
	}
	\label{app_fig:figure1}
\end{figure}
In this section, we study the Hamiltonian of an isolated superconducting cavity (SC), restricted to a strip geometry with a finite length $L$ in the $x$-direction and infinite width in the $y$-direction. The spectrum of the SC drastically changes depending on the type of superconductivity involved, namely $s$- or $p$-wave pairing, which introduces coupling between electron and hole [cf.~Eq.~(3) of the main text]. Neglecting the pairing potential altogether, independent standing modes are formed in the $x$-direction, while in the $y$-direction the wavefunctions are simply free waves. Once $s$-wave coupling is introduced, 
we show that only the standing modes with the same orbital quantum number hybridize, and open the main gap $\Delta_s$ in the energy spectrum. On the other hand, if the coupling is of $p$-wave type, each standing mode in the $x$-direction is coupled to all other modes that have the opposite parity (including itself). Therefore, besides the main gap that opens in the spectrum, additional secondary gaps appear at higher energies, see Fig.~\ref{app_fig:figure1}. This finding explains the behavior of the transmission in Fig.~2 of the main text in the limit of a high barrier ($Z \gg 1$), and consequentially the behavior of the visibility of the oscillations in Fig.~3 of the main text.    

Let us start by rewriting the BdG Hamiltonian of the isolated SC in real space in terms of fermionic field operators as
\begin{align}   \label{app_eq:SC_hamiltonian}
    \mathcal{H}_{\rm SC} &= \int \dd \mathbf{r} \, \Psi^{\dag}(\mathbf{r}) \, H_{\rm SC}(\hat{\mathbf{k}}) \, \Psi(\mathbf{r}),\\
    H_{\rm SC}(\hat{\mathbf{k}})&=
        \begin{pmatrix}
        \frac{-\hbar^2}{2 m_{\rm S}} \mathbf{\nabla}^2 - \mu_{\rm S} & \Delta(-i\nabla)\\
        \Delta^\dag(-i\nabla) & \frac{\hbar^2}{2 m_{\rm S}} \mathbf{\nabla}^2 + \mu_{\rm S}
        \end{pmatrix} \, ,
\end{align}
where $\Psi(\mathbf{r})=(\psi(\mathbf{r}), \psi^\dag(\mathbf{r}))^T$ is the Nambu spinor, and the wave vector is replaced by the operator $\hat{\mathbf{k}}\equiv{-i\mathbf{\nabla}}$. Here, we have dropped the spin index for brevity which is irrelevant to our analysis. Physically, $s$-wave paired electrons are in the spin-singlet states with opposite spins while the $p$-wave pairing occurs either between spinless fermions or between spinful fermions in the spin-triplet states. Here, we only concentrate on the orbital part of the wave function. 
The field operators for the infinite strip geometry can be expanded as   
\begin{align}
    \psi(\mathbf{r})&=\sqrt{\frac{2}{LW}}\sum_{\alpha,k_y}e^{ik_y y}\sin(\frac{\alpha\pi}{L}x)d_{\alpha,k_y},\\
    \psi^{\dag}(\mathbf{r})&=\sqrt{\frac{2}{LW}}\sum_{\beta,k_y'}e^{-ik_y' y}\sin(\frac{\beta\pi}{L}x)d^{\dag}_{\beta,k_y'} \, ,
\end{align}
with $d,d^{\dag}$ the fermionic annihilation and creation operators, and $\alpha,\beta \in \mathbb{N}$ are the quantum numbers associated with the standing modes in the $x$-direction.
We can now make use of the orthogonal normalization conditions
\begin{align}
    \frac{2}{L}\int_0^L \dd x \sin(\frac{\alpha\pi}{L}x)\sin(\frac{\beta\pi}{L}x)&=\delta_{\alpha,\beta}, \nonumber\\
    \frac{1}{W}\int_{-\infty}^{\infty} \dd y \, e^{i(k_y-k_y')y}&=\delta_{k_y,k_y'},
\end{align}
to obtain the Hamiltonians in the second quantization
\begin{align}   \label{app_eq:Hs_d}
     \mathcal{H}_{\rm SC}^s =& \sum_{\alpha,k_y} D^\dag_{\alpha,k_y}
    \begin{pmatrix}
        \varepsilon_{\alpha}(k_y) & \Delta_s\\
        \Delta_s^* & -\varepsilon_{\alpha}(k_y)
        \end{pmatrix} D_{\alpha,k_y} \,,\\
    \mathcal{H}_{\rm SC}^p =& \sum_{\alpha,\beta,k_y} D^\dag_{\alpha,k_y}
    \begin{pmatrix}
        \varepsilon_{\alpha}(k_y)\delta_{\alpha,\beta} & -\Delta_p[k_y\delta_{\alpha,\beta}+q(\alpha,\beta)]\\
        -\Delta_p^*[k_y\delta_{\alpha,\beta}-q(\alpha,\beta)] & -\varepsilon_{\alpha}(k_y)\delta_{\alpha,\beta}
        \end{pmatrix} D_{\beta,k_y} \, , \label{app_eq:Hp_d}
\end{align}
where the fermionic operator is defined as $D_{\alpha / \beta,k_y}=(d_{\alpha / \beta,k_y},d_{\alpha / \beta,-k_y}^\dag)^T$, the superscript $s,p$ denote the $s$- and $p$-wave pairing, respectively, and
\begin{align} \label{app_eq:delta_0_energies}
    \varepsilon_{\alpha}(k_y) &= \frac{\hbar^2k_y^2 +\hbar^2\alpha^2\pi^2/L^2 }{2m_{\rm S}} - \mu_{\rm S} \, , \\
    q(\alpha,\beta) &=\frac{2\alpha\beta}{L(\alpha^2-\beta^2)}[\cos(\pi\alpha)\cos(\pi\beta)-1]. \label{app_eq:qab}
\end{align}
From Eq.~\eqref{app_eq:Hs_d} it follows that in the case of an $s$-wave superconductor, standing modes with different quantum numbers $\alpha$ remain uncoupled and the band structure is simply $E_{\alpha}(k_y) = \sqrt{\varepsilon_{\alpha}^2(k_y) + \Delta_s^2}$, see the lower panel of Fig.~\ref{app_fig:figure1}. The main gap is opened due to the coupling between electron and hole modes with the same $\alpha$.
On the other hand, the situation is more complicated in the $p$-wave superconductor. Besides the opening of the main gap, caused by the off-diagonal term $-\Delta_pk_y$ in Eq.~\eqref{app_eq:Hp_d}, there are secondary gaps opening at crossings of modes with different $\alpha$ quantum numbers due to the additional off-diagonal term $-\Delta_p q(\alpha,\beta)$, see the upper panel of Fig.~\ref{app_fig:figure1}. Furthermore, the factor $\cos(\alpha \pi) \cos(\beta \pi)-1$ in Eq.~\eqref{app_eq:qab} vanishes if $\alpha$ and $\beta$ are of the same parity and it is equal to -2 if the parities are opposite. This explains the double helix shape that bands exhibit at low $k_y$ in the upper panel of Fig.~\ref{app_fig:figure1}. From the prefactor $\alpha \beta / (\alpha^2-\beta^2)$ in Eq.~\eqref{app_eq:qab} it follows that the coupling between different standing modes reduces with increasing difference between the quantum numbers, which results in the reduction of the size of secondary gaps when energy is increased away from the main gap.
As a consequence, the width of the bands resembling a double helix reduces for bands closer to the main gap, and they become flatter. The same effect occurs in the transmission plot for high barriers, see Fig.~2(d) and (f) of the main text. Note that the secondary gaps, being opened by the pairing potential, are robust to the change of coupling with the external metallic leads, which explains the high values for visibility for $Z \gg 1$ in Fig.~3 of the main text.

%
\subsection{Introducing the coupling to leads as a small perturbation  \label{app_sec:coupling_to_leads}}
%
In Sec.~\ref{app_sec:isolated_cavity}, we showed how the momentum-dependence of the $p$-wave superconducting pairing opens the secondary gaps in the band structure of an SC that is finite in the $x$-direction and infinite in the $y$-direction. On the other hand, the $s$-wave coupling does not hybridize different standing modes of the SC, and the secondary gaps remain closed for any value of $\Delta_s$. 
In this section, we introduce weak tunnel-coupling $t$ between the SC and the metallic leads, i.e., high, but finite barrier $Z > 1$. Thus, we show that hybridization with the metals can open secondary gaps, which explains the structure of transmission in Fig.2(a) -- and with that the existence of oscillations in conductance -- for the $s$-wave superconductor.

Here, we concentrate on the $s$-wave SC when it is coupled to two-dimensional metallic leads that are infinite in the $y$-direction and semi-infinite in the $x$-direction. In the following, we treat the tunneling Hamiltonian
as a small perturbation and obtain an effective Hamiltonian for the SC.
The BdG Hamiltonian of a whole system is 
\begin{equation}
    \mathcal{H} = \mathcal{H}^s_{\rm SC} + \mathcal{H}_{\rm N} + \mathcal{H}_{\rm T}\,,
\end{equation}
where $\mathcal{H}^s_{\rm SC}$ is given in Eq.~\eqref{app_eq:Hs_d}, the metallic leads are described by 
\begin{align}
    \mathcal{H}_{\rm N} &= \sum_{\mathbf{k}}C_{\mathbf{k}}^{\dagger} H_{\rm N}(\mathbf{k}) C_{\textbf{k}} \,,
\end{align}
with $H_{\rm N}(\mathbf{k})$ given by Eq.~(2) of the main text
and $C_{\textbf{k}}=(c_{\textbf{k}}, c_{-\textbf{k}}^\dag)$ is the fermionic operator
in the leads,
and the tunnel coupling term reads
\begin{align}
    \mathcal{H}_{\rm T} &= \sum_{\textbf{k},\alpha}C^\dag_{\textbf{k}}H_{\rm T}D_{\alpha,k_y} + h.c. \, ,\\
    H_{\rm T}&=\begin{pmatrix}
        t & 0\\
        0 & -t^*
        \end{pmatrix},
\end{align}
with the coupling strength being $t$ which, for the moment, we assume to be constant. 

Applying the unitary transformation,
\begin{align}                                                                                
d_{\alpha,k_y} &= u_\alpha\gamma_{\alpha,k_y}+v_\alpha\gamma_{\alpha,-k_y}^{\dagger} \, ,
\end{align}
with 
\begin{align}
    u_\alpha^2=1-v_\alpha^2=\frac{1}{2}\left(1+\frac{\varepsilon_\alpha(k_y)}{E_\alpha(k_y)}\right) \,  ,
\end{align}
and 
\begin{align}
    E_\alpha(k_y)=\sqrt{\varepsilon_\alpha(k_y)^2+|\Delta_s|^2}\,,
\end{align}
we can diagonalize $\mathcal{H}^s_{\rm SC}$ [cf.~Eq.~\eqref{app_eq:Hs_d}] as
\begin{align}\label{B8}
    \mathcal{H}^s_{\rm SC} =& \sum_{\alpha,k_y}  \Gamma_{\alpha,k_y}^{\dagger}\begin{pmatrix}
        E_{\alpha}(k_y) & 0\\
        0 & -E_{\alpha}(k_y)
        \end{pmatrix}\Gamma_{\alpha,k_y} \, , 
\end{align}
where $\Gamma_{\alpha,k_y}=(\gamma_{\alpha,k_y},\gamma^\dag_{\alpha,-k_y})^T$.
Note that $\mathcal{H}^s_{\rm SC}$ is diagonal in the basis of the standing modes inside the superconducting region.
At the same time, the tunneling Hamiltonian is transformed into
\begin{align}
\mathcal{H}_{\rm T} =& \sum_{\textbf{k}, \alpha}  C_{\textbf{k}}^{\dagger}\tilde{H}^\alpha_{\rm T}\Gamma_{\alpha,k_y}+h.c. ,\\
\tilde{H}^\alpha_{\rm T}=&\begin{pmatrix}
        u_\alpha(k_y)t & v_\alpha(k_y)t\\
        -v_\alpha(k_y)t^* & -u_\alpha(k_y)t^*
        \end{pmatrix}.
\end{align}

\begin{figure}[t!]
    \centering
    \includegraphics[width=0.6\columnwidth]{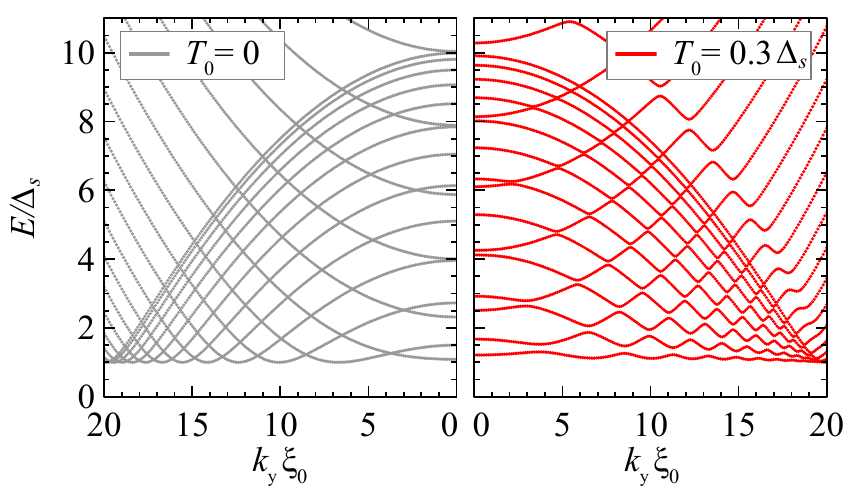}
    \caption{
    Band structure of an $s$-wave SC [infinite in the $y$-direction, and finite in the $x$-direction with length $L$]. (left plot) No coupling to the leads. (right plot) Coupling to the leads lifts the degeneracy of all crossings and opens secondary gaps in the spectrum also away from the main gap. To emphasize all the features of the band structure, i.e. shape of the sombrero hat, the main gap, regimes inside/outside the sombrero hat where only particles/both particles and holes exist and secondary gaps opening when coupling to the leads is included, we changed the parameters -- compared to the main text -- and used the following: $L=2 \xi_0$, $k_{\rm F}=20 \xi_0^{-1}$, $\mu_{S}=10 \Delta_s$, where $\xi_0$ is the superconducting coherence length.
    }
    \label{app_fig:figure2}
\end{figure}
Let us now calculate the self-energy correction to the SC Hamiltonian that arises because of the coupling with the leads. We choose 
$\left( \Gamma_{\{\alpha=1,\cdots,n\},k_y}, C_{\{k_x\in(-\infty,\infty)\},k_y}\right)$
as the basis and the matrix elements of the self-energy of the SC due to the tunnel coupling is calculated by
\begin{align}
    \Sigma_{\alpha\beta}(\omega, k_y)&=\sum_{k_x}\tilde{H}_{\rm T}^{\alpha\dag}(k_y)G_{\rm N}(\textbf{k},\omega)\tilde{H}_{\rm T}^{\beta}(k_y),\\
    G_{\rm N}(\textbf{k},\omega)&=\frac{1}{\omega-\epsilon(\textbf{k})\sigma_z+i0^+},
\end{align}
where $G_{\rm N}(\textbf{k},\omega)$
is the retarded Green's function for the leads with $\epsilon(\textbf{k})=\hbar^2(k_x^2+k_y^2)/(2m_{\rm N})-\mu_{\rm N}$ the energy of electron therein.

We are interested in the opening of the secondary gaps between different standing modes in the SC. The positive and negative eigenvalues $\{\pm E_\alpha(k_y)\}$ have already been separated by the main gap $\Delta_s$ so that it is sufficient to consider the matrix elements of the self-energy that couple different modes with positive energy $\{+E_\alpha\}$, which reads
\begin{align}
    \Sigma_{\alpha\beta}^{++}(k_y)&=T_{\alpha\beta}(k_y)=T_0(u_\alpha u_\beta+v_\alpha v_\beta), \nonumber \\
    T_0&=\sum_{k_x}\frac{|t|^2}{\omega-\epsilon(\textbf{k})+i0^+}.
\end{align}
Moreover, given that the high interface barrier has pushed the electronic states in the leads (in the vicinity of the interface) far away above the
gap $\Delta_s$, $T_0$ can be assumed to be real and constant in the interested energy interval $\omega\sim \Delta_s$.
We can now write down the effective Hamiltonian which describes the bands of the SC with positive energy as 
\begin{align}
    \mathcal{H}^{s+}_{\rm SC} =& \sum_{\alpha,\beta,k_y}  \gamma_{\alpha,k_y}^{\dagger}
       \left[E_{\alpha}(k_y) \delta_{\alpha,\beta}+T_{\alpha\beta}(k_y)\right] \gamma_{\beta,k_y},
\end{align}
or in its matrix form as
\begin{align}\label{app_eq:effective_hamiltonian}
    H^{s+}_{\rm SC} =\left(
    \begin{array}{cccc}
    E_1+T_{11}&T_{12}&\cdots&T_{1n}\\
   T_{21}& E_2+T_{22}& \cdots&\vdots\\
   \vdots& \vdots & \ddots & T_{n-1,n}\\
    T_{n1}& \cdots & T_{n,n-1}&E_n+T_{nn}
    \end{array}\right)_{n\times n} \, ,
\end{align}
where the truncation is made to the lowest $n$ standing modes with the energy much higher than $\Delta_s$.
One can see that the self-energy introduces the off-diagonal elements that couple different standing modes
and thus open secondary gaps between neighboring standing modes inside
the cavity. We plot the energy modes of Eq.~\eqref{app_eq:effective_hamiltonian} numerically in Fig.~\ref{app_fig:figure2}.

%
\subsection{Conductance as a function of the length of the superconducting cavity  \label{app_sec:different_length}}
%
\begin{figure}[h!]
    \centering
    \includegraphics[width=\textwidth]{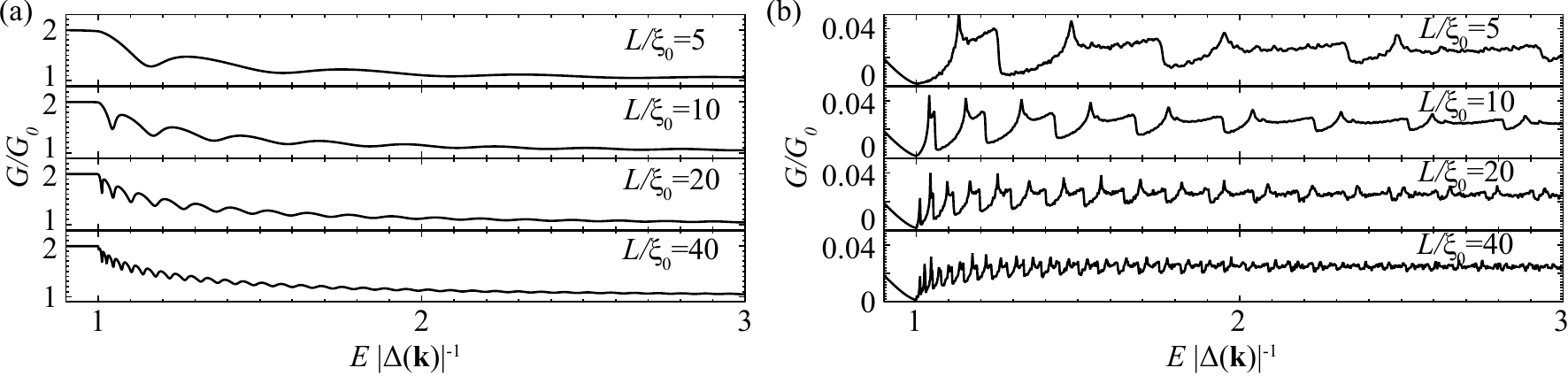}
    \caption{
    Differential conductance as a function of energy through the NSN junction containing a $p$-wave superconductor for different lengths $L$ and for (a) $Z=0$ and (b) $Z=4$.
    }
    \label{app_fig:different_length}
\end{figure}
In Fig.~\ref{app_fig:different_length}, we show differential conductance obtained from Eq.~\eqref{eq:conductance} for the case of a $p$-wave superconducting cavity for different cavity lengths. In Fig.~\ref{app_fig:different_length}(a), a transparent limit with $Z=0$ is presented, while in Fig.~\ref{app_fig:different_length}(b) we show the tunnelling limit with $Z=4$. For both weak and strong barriers, the energy dependent period of oscillations doubles when the length $L$ is halved, as it is expected from the interference condition $k_x^{e/h}(E)-k_x^{h/e}(E) = 2 \pi n / L$, where $n$ is an integer. More details about this limit can be found in Refs.~\cite{Karalic2020,thesis_Strkalj}.

Furthermore, in both cases the height of oscillations compared to the background, and therefore their visibility, reduces with increasing $L$. For the transparent limit shown in Fig.~\ref{app_fig:different_length}(a) this effect can be contributed to the broadening of the resonances in the superconducting cavity caused by the hybridization with the external leads, similarly to the case of 1D Fabry-P\'{e}rot oscillations. Bringing the resonances closer together by increasing $L$ results in increase of the background conductance in the energy interval between the resonances.  

On the other hand, in the tunnelling limit, shown in the Fig.~\ref{app_fig:different_length}(b), besides the small broadening of the resonances caused by a weak coupling to the leads, the height of oscillations additionally reduces because of the fact that the off-diagonal elements in Eq.~\eqref{app_eq:Hp_d}, which are responsible for opening the secondary gaps, scale as $1/L$, see Eq.~\eqref{app_eq:qab}.

%
\subsection{A short survey of the advantages and drawbacks of contemporary methods proposed for the identification of topological superconductors  \label{app_sec:advanteges_drawbacks}}
%

Heretofore, several methods have been proposed with the goal of unequivocally detecting topological superconductivity. These detection schemes are based on unique signatures that these systems should have, such as surface states, zero-bias peak in vortex cores where the Majorana mode is expected to exist, and the quantisation of the aforementioned states. All of these in-gap signatures, unfortunately, can emerge from other trivial reasons, e.g., from disorder that can introduce trivial surface states and result in a zero-bias peak, Caroli-de Gennes-Matricon states in vortices~\cite{Caroli1964,Prada2020}, and Yu-Shiba-Rusinov states~\cite{Yu1965,Shiba1968,Rusinov1969}, that can generate similar experimental signatures as those of Majorana modes. 
Our proposal is immune to the aforementioned in-gap disorder-induced effects, since it concentrates on interference effects above the gap. The same disorder, that would produce false-positive results in other measurements, will close the secondary gaps above the bottom of the band, and reduce the oscillations in conductance -- therefore giving a correct, negative result in this case. 

Other phase-sensitive probes that provide some information about the gap symmetry make use of cleverly designed SQUID interferometers, e.g., corner~\cite{VanHarlingen1995} or GLB SQUID interferometers~\cite{Nelson2004} attached to 3D superconducting crystals. However, such experiments are notoriously difficult to perform and include many effects induced by contact resistance between two different types of superconductors. Moreover, they can discern the parity of the gap phase, but cannot resolve the $p$- versus $f$-wave issue~\cite{Mackenzie2003}. Furthermore, the aforementioned experiments are relevant for 3D superconducting samples, but their generalisation to 2D systems is still lacking.

Lastly, let us mention some clear advantages and disadvantages of our method. In contrast to the common in-gap measurements that can give false-positive results caused by trivial effects of disorder, the Tomasch oscillations, being a coherent phenomenon, would be washed out in the presence of disorder and, therefore, correctly identify the trivial superconductor. 
However, in the case of a chiral $d$-wave superconductivity -- where the pairing is an even function of momenta -- the secondary gaps will remain closed, and therefore, we do not expect to observe any oscillations in conductance. This would clearly be a false negative result of our method.


\begin{thebibliography}{71}
\bibitem [{foo()}]{footnote_basis}%
	\BibitemOpen
	\href@noop {} {}\bibinfo {note} {Note that the Hamiltonian in the $s$-wave
		case is written in the basis $(d^{\phantom\dagger}_{k,\sigma},
		d^\dagger_{-k,-\sigma})^T$ with $\sigma=(\uparrow, \downarrow)$, and the one
		for the $p$-wave case has a spinless basis $(d_{k}, d^\dagger_{-k})^T$, where
		$d$ and $d^\dagger$ are the electron annihilation and creation operator,
		respectively.}\BibitemShut {Stop}%
\end{thebibliography}
\end{document}